\pdfoutput=1
\documentclass[lettersize,journal]{IEEEtran}

\usepackage{amsmath,amsfonts,amssymb}
\usepackage{algorithmic}
\usepackage{algorithm}
\usepackage{array}
\usepackage{textcomp}
\usepackage{stfloats}
\usepackage{url}
\usepackage{graphicx}
\usepackage{cite}
\usepackage{bm}
\usepackage{booktabs}
\usepackage[T1]{fontenc}
\usepackage{newtxmath}

\begin{document}

\title{SafeSABR: Risk-Calibrated Adaptive Bitrate Streaming over Starlink Networks}
\author{Hongjun Xie, Jiahang Zhu, Zhiming Shao, Chao Fan, Zenghui Zhang, Genke Yang, and Pengcheng Luo\thanks{This work was supported by the National Major Science and Technology Project for Intelligent Manufacturing Systems and Robotics of China under Grant 2025ZD1602400. \emph{(Corresponding author: Pengcheng Luo.)}}
\thanks{Hongjun Xie, Chao Fan, Zenghui Zhang, Genke Yang and Pengcheng Luo are with Ningbo Artificial Intelligence Institute, Shanghai Jiao Tong University, Ningbo 315000, China, and also with the School of Automation and Intelligent Sensing, Shanghai Jiao Tong University, Shanghai 200240, China, and the Key Laboratory of System Control and Information Processing, Ministry of Education of China, Shanghai 200240, China (e-mail: xiehongjun@sjtu.edu.cn, fchao2025@sjtu.edu.cn, zenghui.zhang@sjtu.edu.cn, gkyang@sjtu.edu.cn, luopeng69131@sjtu.edu.cn).}
\thanks{Hongjun Xie and Pengcheng Luo are also with Shanghai i-Space Orbital Computing Infrastructure Technology Co., Ltd., Shanghai 200235, China.}
\thanks{Jiahang Zhu is with Ningbo Industrial Internet Institute, Ningbo 315000, China (e-mail: zhujiahang2018@163.com).}
\thanks{Zhiming Shao is with the School of Automation and Intelligent Sensing, Shanghai Jiao Tong University, Shanghai 200240, China (e-mail: zm.shao@sjtu.edu.cn).}
\thanks{The source code of SafeSABR is available at: \protect\url{https://github.com/luopeng69131/SafeSABR}.}
}

\maketitle

\begin{abstract}
Starlink, as a representative low Earth orbit (LEO) satellite broadband system, makes high-bitrate video streaming possible in regions where terrestrial broadband is unavailable. However, its access links exhibit rapid throughput fluctuations caused by satellite mobility and handovers. Existing learned adaptive bitrate (ABR) algorithms can achieve high average quality of experience (QoE), yet high-bitrate Starlink streaming exposes severe session-level rebuffering that is not captured by average QoE alone. To address it, this paper proposes \emph{SafeSABR}, a risk-calibrated learned ABR framework for Starlink networks. SafeSABR formulates Starlink ABR as a QoE--severe-risk tradeoff and follows a three-stage design: behavior-cloning pretraining learns a high-QoE ABR prior, risk-calibrated reinforcement learning (RL) fine-tuning reduces severe-tail action tendencies, and a runtime safety auditor uses safe-capacity lower bounds to check policy-requested bitrates before execution. Experiments on real Starlink traces compare SafeSABR with online, prediction-assisted, and learned ABR baselines. Compared with advanced methods, SafeSABR reduces severe-stall sessions from 22.8\% to 7.2\% and worst-5\% session rebuffering from 54.30 s to 22.68 s, with a 1.8\% QoE cost. Component analyses further show that risk-calibrated fine-tuning and safe-capacity auditing reduce unsafe bitrate decisions and downstream severe-session rebuffering. These results show that combining risk-calibrated policy learning with decision-aware safe throughput forecasting can move learned ABR toward a safer QoE--severe-risk operating point under volatile Starlink networks.
\end{abstract}


\section{Introduction}
\subsection{Background and Motivation}
\begin{figure}[!t]
  \centering
  \includegraphics[width=\linewidth]{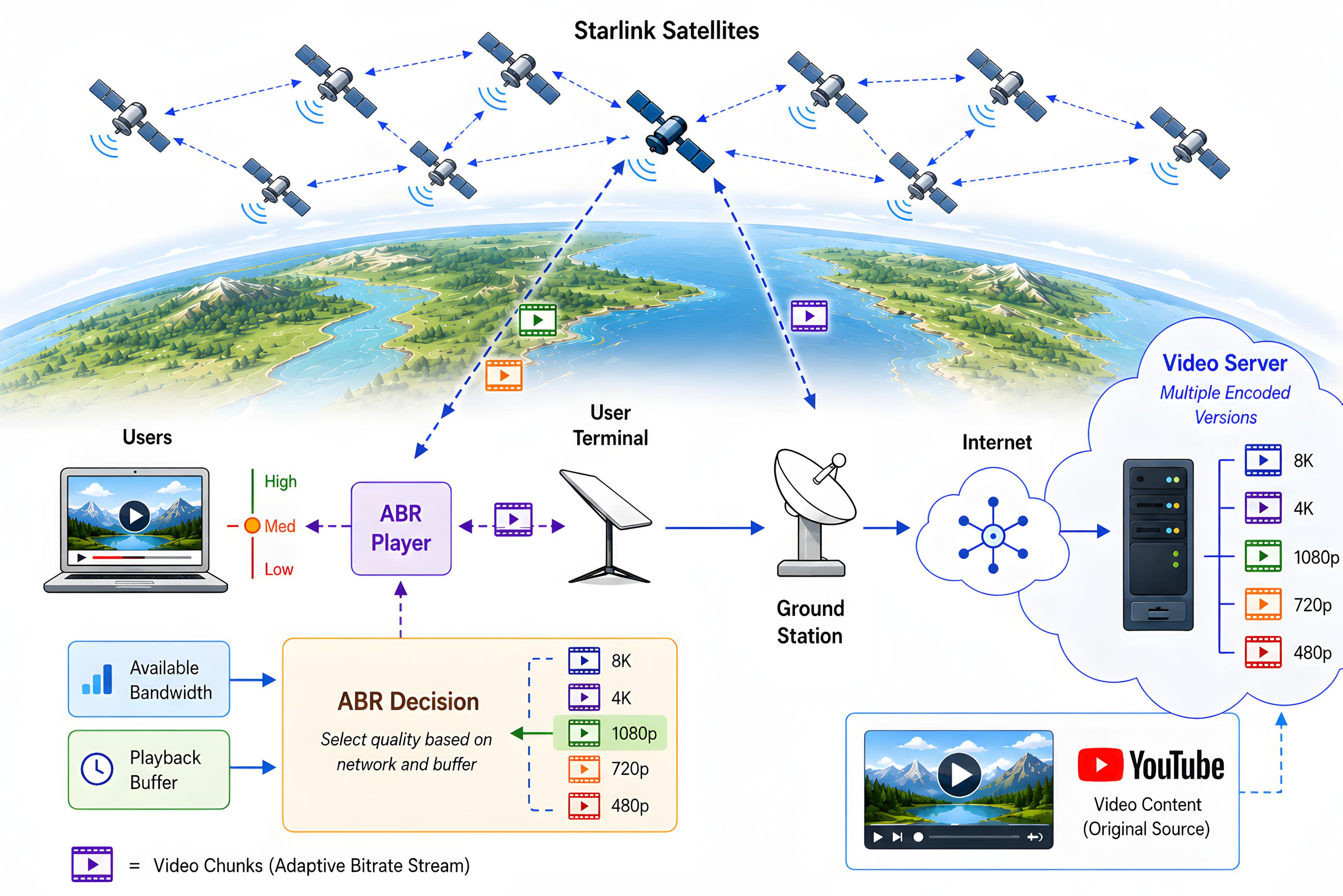}
  \caption{Illustrative ABR-over-Starlink video delivery scenario. The video server stores multiple encoded versions of the same content, such as 480p, 720p, 1080p, 4K, and 8K representations. During playback, the ABR player observes the available bandwidth and playback buffer, selects the next chunk representation, and requests the selected chunk over the Starlink access path through the user terminal, satellite network, gateway, and Internet.}
  \label{fig:intro-background}
\end{figure}

\IEEEPARstart{A}{daptive} bitrate (ABR) streaming is a representative bandwidth-demanding application enabled by Starlink broadband access. Compared with traditional satellite Internet, Starlink-like low Earth orbit (LEO) systems can provide higher access capacity and lower latency, making high-bitrate video streaming feasible in rural areas, oceans, airborne platforms, disaster recovery scenarios, and other regions where terrestrial broadband is unavailable~\cite{michel2022first,kassem2022browser,TWC23CKJ}. In such scenarios, a video service stores multiple encoded versions of the same content, and an ABR client continuously selects the representation of the next video chunk according to the playback buffer and the expected future throughput. A higher-quality representation improves video quality when the link remains strong, but it can quickly lead to rebuffering when the available capacity drops. Fig.~\ref{fig:intro-background} illustrates this ABR-over-Starlink delivery pipeline and the client-side resolution/bitrate selection process.

\begin{figure*}[!t]
  \centering
  \includegraphics[width=0.85\textwidth]{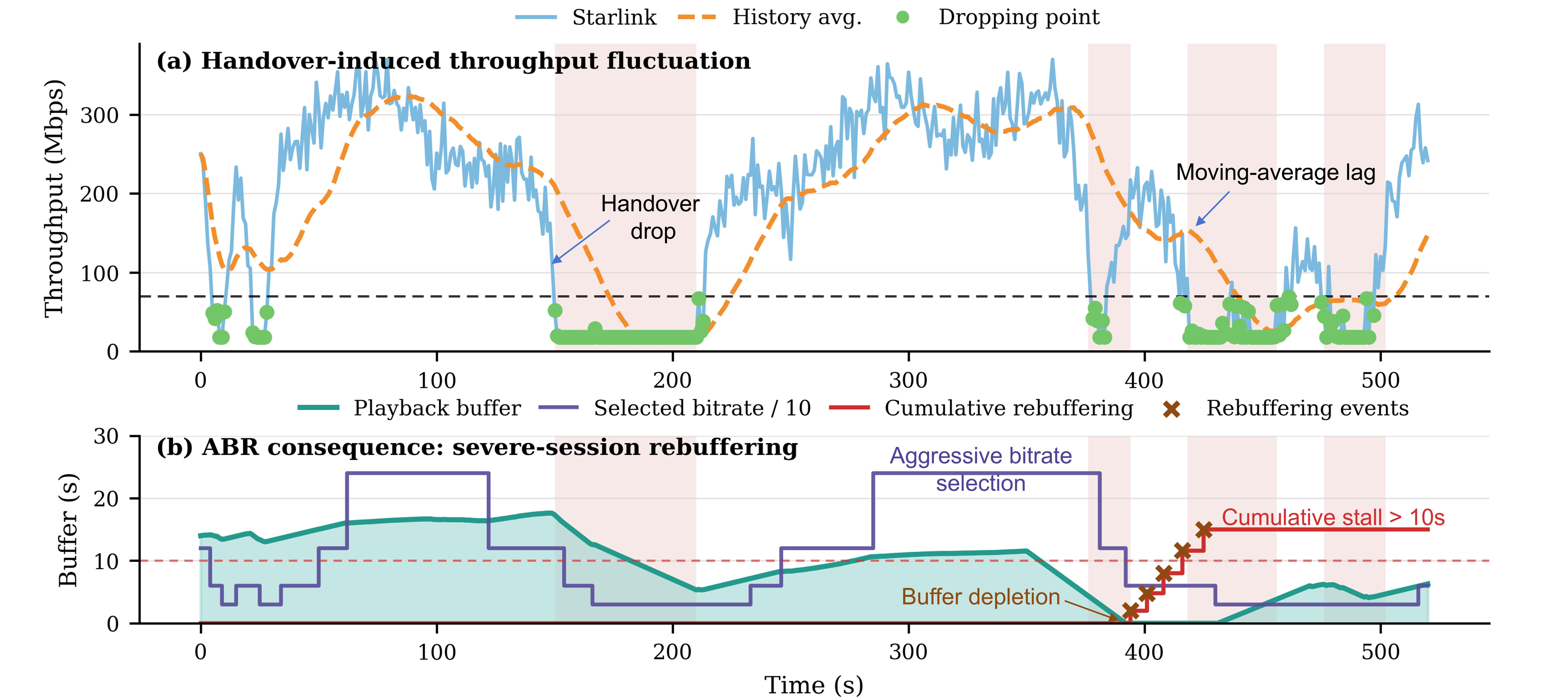}
  \caption{Challenge of ABR streaming over volatile Starlink access links. Handover-induced throughput drops and history-average lag can make an ABR client maintain an aggressive bitrate after the actual link capacity has decreased. The selected bitrate then drains the playback buffer, creates multiple rebuffering events, and accumulates more than 10 s of session-level stall.}
  \label{fig:intro-challenge}
\end{figure*}

Starlink access throughput is highly dynamic~\cite{garcia2023multi,Tcom25CKJ}. Measurement studies have shown that user-perceived throughput can fluctuate over short time scales due to satellite handovers, elevation-angle changes, obstruction, gateway association, weather, and traffic load~\cite{liu2025starnet}. This volatility changes the nature of ABR control. In terrestrial broadband or cellular traces, a high recent throughput often provides a useful signal for selecting a higher bitrate~\cite{tiwari2023t3p,CM25CKJ}. In high-bitrate Starlink streaming, however, the same history can become misleading: an ABR client may keep requesting a large chunk just before the access link drops. The resulting problem is therefore not only how to increase average video quality, but also how to avoid session-level severe rebuffering caused by abrupt throughput collapses, as illustrated in Fig.~\ref{fig:intro-challenge}. This motivates a Quality of Experience (QoE)--severe-risk view of Starlink ABR, where a method should be judged not only by mean QoE but also by the worst-session rebuffering tail and the fraction of sessions with unacceptable cumulative stalls.

Existing learning-based ABR research mainly pursues higher QoE by improving policy learning, adaptation, and generalization across network conditions. Genet uses automatic curriculum generation to expose adaptation policies to diverse network conditions~\cite{xia2022genet}. Offline RL and meta-RL studies further improve bitrate adaptation across heterogeneous traces and tasks~\cite{yi2025optimizing,bentaleb2023meta}, while bitrate-guidance, meta-learning, and information-theoretic neural adaptation methods improve cross-condition generalization~\cite{bentaleb2024bitrate,li2023metaabr,kan2025merina+}. Emerging large-model-based networking studies also suggest new opportunities for context-aware control and network adaptation~\cite{wu2024netllm}. These methods show the potential of learned control for improving ABR performance, but their optimization and evaluation are still largely driven by high QoE or average performance. In highly volatile Starlink access links, handover-induced throughput drops can make a learned ABR policy keep requesting aggressive high-bitrate chunks, and a small number of such decisions may drain the playback buffer and produce severe session-level stalls. The remaining gap is therefore not simply better ABR adaptation, but a risk-calibrated ABR design that preserves high QoE while explicitly reducing severe-stall risk caused by abrupt Starlink throughput drops.

Reducing this severe-stall risk requires connecting throughput prediction with the bitrate action that will be executed. A throughput overestimation may be harmless when the playback buffer is large or the selected chunk is small, but the same overestimation can trigger rebuffering when the buffer is low and the requested chunk is large. Therefore, evaluating a predictor only by point-estimation accuracy is insufficient for Starlink ABR~\cite{xie2026riskawaresafethroughputforecasting}. The policy must be calibrated toward lower severe-tail risk rather than only maximizing mean QoE, and the predictor must expose a safe-capacity estimate that can be checked against the actual bitrate action being requested.

Motivated by this observation, this paper proposes \emph{SafeSABR}, a risk-calibrated learned ABR framework for Starlink streaming with safe-capacity-audited runtime decisions. Following the pretraining--fine-tuning framework of SABR~\cite{luo2025sabr}, SafeSABR first uses behavior-cloning pretraining to obtain a high-QoE ABR prior and then performs risk-calibrated RL fine-tuning, instantiated by conditional value-at-risk proximal policy optimization (CVaR-PPO), to reduce severe-tail action tendencies. At deployment time, a runtime safety auditor uses a safe-capacity estimate to check and correct high-risk bitrate requests before execution. We instantiate the safe-capacity input with BG-CFQS~\cite{xie2026riskawaresafethroughputforecasting}, which forecasts Starlink throughput lower bounds for risk-aware control. The design follows a QoE-oriented pretraining, risk-calibrated fine-tuning, and safety-audited deployment pipeline.

\begin{figure*}[!t]
  \centering
  \includegraphics[width=0.98\linewidth]{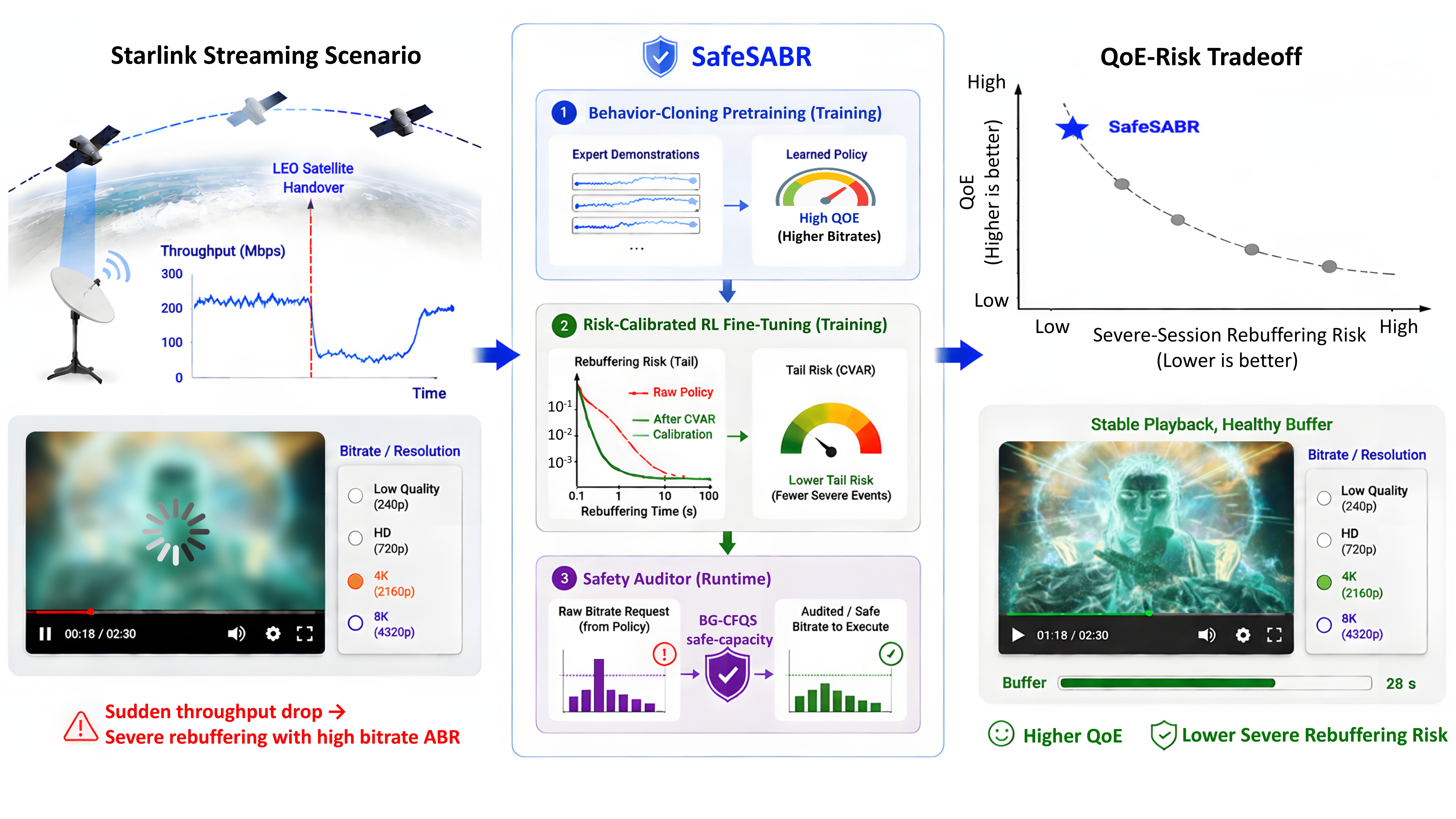}
  \caption{Overview of SafeSABR. SafeSABR addresses the high-bitrate Starlink ABR problem by learning a high-QoE prior through behavior-cloning pretraining, applying risk-calibrated RL fine-tuning with CVaR-PPO, and auditing unsafe bitrate requests with BG-CFQS safe-capacity forecasting at runtime.}
  \label{fig:intro-safesabr-concept}
\end{figure*}

\subsection{Contributions}
The main contributions of this paper are summarized as follows:
\begin{itemize} 
\item We formulate high-bitrate Starlink ABR as a QoE--severe-risk tradeoff problem rather than an average-QoE ranking problem. In addition to average QoE, we use session-level cumulative rebuffering metrics, including worst-5\% session rebuffering and the fraction of sessions with more than 10s rebuffering, to expose severe playback failures hidden by mean performance.

\item We develop risk-calibrated policy learning in SafeSABR by extending the SABR pretraining--fine-tuning paradigm from average-QoE learning to severe-risk-aware ABR control. Behavior cloning provides a high-QoE ABR prior, while risk-calibrated RL fine-tuning, instantiated by CVaR-PPO, reshapes the learned policy toward lower severe-tail risk.

\item We design a decision-aware runtime safety auditor driven by safe throughput forecasting. The auditor converts BG-CFQS safe-capacity lower bounds into ABR action-level feasibility checks, and the predictor-auditor interface is evaluated by decision violation, high-risk overestimation, audit rate, and downstream severe-session rebuffering.

\item We conduct a comprehensive evaluation on real-world Starlink throughput traces, including comparisons with representative ABR baselines and analyses of robustness, components, and runtime mechanisms. The results show that SafeSABR reduces severe-stall sessions from 22.8\% to 7.2\% with a 1.8\% QoE cost, and achieves a more favorable QoE--severe-risk operating point among high-QoE methods.
\end{itemize}

\subsection{Organization}
The remainder of this paper is organized as follows. Section II reviews related work on LEO satellite measurement and video streaming, Starlink throughput prediction, and ABR algorithms. Section III formulates the high-bitrate Starlink ABR problem, including the chunk-level decision model, QoE--severe-risk metrics, and safe-capacity action feasibility. Section IV presents the SafeSABR framework, including behavior-cloning pretraining, risk-calibrated RL fine-tuning, decision-aware safe-capacity prediction, and runtime safety auditing. Section V describes the experimental setting and evaluates SafeSABR through main comparisons, robustness tests, mechanism analysis, predictor-auditor analysis, ablation, and sensitivity studies. Section VI concludes the paper.

\section{Related Work}

The related literature is organized along three lines. We first review Starlink measurement and video streaming over LEO, which characterize the access environment considered in this paper. We then discuss Starlink throughput prediction, which is closely related to bitrate decision making over volatile links. Finally, we review ABR algorithms, including classical online methods and recent learning-based approaches.

\subsection{Starlink Measurement and Video Streaming over LEO}

The rapid deployment of Starlink has motivated extensive measurement studies on LEO satellite broadband. Early active and browser-side measurements characterized Starlink throughput, latency, and packet loss, showing that commercial LEO access can provide broadband-level capacity but with noticeable temporal variability~\cite{michel2022first,kassem2022browser}. End-user and multi-terminal studies further reported that performance varies across location, time, terminal status, and application workload~\cite{ma2023network,mohan2024multifaceted}. These results establish Starlink as a viable access network for high-bitrate applications, but also indicate that its dynamics differ from conventional terrestrial broadband.

Several studies have further examined the causes and application impact of such dynamics. Multi-timescale throughput measurements reported both short-term fluctuations and longer-term load patterns~\cite{garcia2023multi}. Other works analyzed the role of scheduling, obstruction, satellite visibility, handovers, and bent-pipe routing in shaping user-perceived performance~\cite{tanveer2023making,tiwari2023t3p,liu2025starnet}. For video applications, prior studies have evaluated real-time multimedia services, Dynamic Adaptive Streaming over HTTP (DASH) streaming, and large-scale video behavior over Starlink~\cite{zhao2023realtime,zhao2024dash,izhikevich2024global}. These works motivate ABR over LEO networks, but they mainly characterize network or application behavior. They do not define a QoE--severe-risk ABR objective or a learned ABR design that explicitly controls session-level rebuffering tails under Starlink throughput drops.

\subsection{Starlink Throughput Prediction}

Throughput prediction is a natural tool for ABR over dynamic access links. For mobile adaptive streaming, Lumos shows that decision-tree throughput prediction can be integrated with ABR control to improve QoE~\cite{lv2022lumos,lv2024accurate}. For Starlink, recent predictors exploit information beyond a short history of measured throughput. T3P uses terminal and satellite-context information to improve LEO throughput prediction~\cite{tiwari2023t3p}. StarNet further incorporates satellite-domain knowledge and handover-aware temporal patterns for Starlink throughput modeling~\cite{liu2025starnet}. Fine-grained burst characterization has also been used to model and predict Starlink throughput variations~\cite{garcia2025modeling}. BG-CFQS further studies risk-aware safe throughput forecasting for Starlink networks and provides calibrated lower-bound capacity estimates for safety-sensitive control~\cite{xie2026riskawaresafethroughputforecasting}. These studies show that throughput can be predicted more effectively when network-specific structure is considered.

Throughput prediction accuracy, however, is not equivalent to ABR decision safety. A point predictor with low average error may still be harmful if it overestimates capacity when the playback buffer is low or when the selected chunk is large. Conversely, a conservative estimate may reduce rebuffering but unnecessarily sacrifice video quality. The relevant question is not only whether a predictor is accurate, but whether its output can be used by an ABR controller to reduce unsafe bitrate decisions and severe-session rebuffering tails.

\subsection{Adaptive Bitrate Streaming}

ABR algorithms select the bitrate of each video chunk to balance video quality, bitrate smoothness, and rebuffering. Classical online methods rely on throughput prediction or buffer occupancy, such as Model Predictive Control (MPC)-based planning and BOLA-style buffer control~\cite{yin2015control,spiteri2020bola}. Learning-based ABR methods instead train bitrate policies from QoE feedback. Beyond early neural and imitation-learning methods such as Pensieve and Comyco~\cite{mao2017neural,huang2019comyco}, recent studies have focused on making learned ABR policies adapt better across diverse network conditions. Genet uses automatic curriculum generation for learning adaptation policies~\cite{xia2022genet}. Offline RL and meta-RL further improve bitrate decisions across heterogeneous traces and tasks~\cite{yi2025optimizing,bentaleb2023meta}, while bitrate guidance, MetaABR, and Merina+ improve cross-condition generalization through meta-learning or information-theoretic neural adaptation~\cite{bentaleb2024bitrate,li2023metaabr,kan2025merina+}. Large-model-based networking systems such as NetLLM also indicate a broader trend toward context-aware network control~\cite{wu2024netllm}.

These studies demonstrate the promise of learned ABR for improving QoE and adaptation performance, but they do not fully address the severe-stall risk created by highly volatile satellite access links. Rebuffering is usually included in the QoE reward, yet expected-QoE optimization can still average out rare but long stalls. General risk-aware optimization and safe RL offer tools for tail-risk learning and runtime action filtering~\cite{rockafellar2000optimization,chow2015risksensitive,chow2018riskconstrained,alshiekh2018safe}, but Starlink ABR safety must be tied to chunk size, requested bitrate, playback buffer, and predicted safe capacity. Therefore, this paper focuses on a complementary problem: how to preserve the high-QoE behavior learned by modern ABR methods while explicitly reducing session-level severe rebuffering under Starlink throughput volatility.

\section{Starlink ABR Model and QoE--Severe-Risk Objective}
\label{sec:formulation}

This section formalizes the two decision links used throughout the paper. First, a bitrate action determines chunk download time, buffer evolution, and rebuffering, which motivates a QoE--severe-risk objective at the session level. Second, a predicted safe capacity determines whether a requested bitrate is feasible before execution, which connects throughput forecasting to ABR action safety. For ease of reading, the main notation is summarized in Table~\ref{tab:notation}.

\begin{table}[!t]
\centering
\caption{Main notation used in the formulation and design.}
\label{tab:notation}
\begin{tabular}{lp{0.68\linewidth}}
\toprule
\textbf{Symbol} & \textbf{Meaning} \\
\midrule
$t$, $T$, $\Delta$ & Chunk index, number of chunks in a session, and chunk duration. \\
$s_t$, $h_t$ & ABR policy state and predictor input at chunk $t$. \\
$b_t$, $B_{\max}$ & Playback buffer and maximum buffer size. \\
$\mathcal{A}$, $a_t$ & Ordered bitrate-action set and selected action. \\
$a_{\min}$, $\mathcal{C}_t$ & Lowest bitrate action and feasible candidate set no higher than the raw request. \\
$r(a)$, $S_t(a)$ & Bitrate and chunk size under action $a$. \\
$c_t$, $\hat c_t$ & Realized throughput and predicted safe capacity. \\
$d_t(a)$, $\rho_t$ & Download time and rebuffering under action $a$. \\
$\mu$, $\eta$ & Rebuffering and bitrate-smoothness penalty weights. \\
$q_t$, $Q$, $R$ & Per-chunk QoE, session-level QoE, and session-level rebuffering. \\
$\tau$, $\pi$ & Streaming session and ABR policy. \\
$\mathcal{J}_Q$, $\mathcal{R}_{\beta}$, $\mathcal{S}_{\rho_0}$ & Expected QoE, tail-rebuffering risk, and severe-stall probability. \\
$\beta$, $\rho_0$, $\mathrm{VaR}$ & Tail-risk confidence level, severe-stall threshold, and value-at-risk. \\
$g$, $\mathcal{F}_t$ & Buffer guard margin and predicted feasible action set. \\
$\alpha$, $\xi_\alpha$, $\lambda$ & CVaR confidence level, empirical tail threshold, and risk penalty weight. \\
$\mathcal{D}_{\mathrm{IL}}$, $a_t^E$, $y_t(a)$ & Imitation dataset, expert action label, and one-hot expert label. \\
$\mathcal{L}_{\mathrm{IL}}$, $\mathcal{B}$, $R_i$ & Imitation loss, rollout batch, and episode-level rebuffering in a batch. \\
$a_t^{\mathrm{raw}}$, $a_t^{\mathrm{safe}}$ & Raw policy action and audited action. \\
$v_t(a)$, $m_t$ & Decision-violation indicator and audit-intervention indicator. \\
$\pi_{\theta}$, $\pi_E$, $\phi$, $\Phi$ & Learned policy, expert policy, safe-capacity predictor, and predictor candidate set. \\
\bottomrule
\end{tabular}
\end{table}

\subsection{Chunk-Level ABR Decision}
We characterize the chunk-level relationship among bitrate selection, download time, buffer evolution, and rebuffering. We consider video streaming over a Starlink access link. A video is divided into chunks indexed by $t$, and each chunk has duration $\Delta$. Before downloading chunk $t$, the ABR controller observes a state $s_t$ that includes the playback buffer $b_t$, recent throughput observations, previous bitrate, remaining chunks, and chunk-size information~\cite{mao2017neural,luo2025sabr}. The controller selects a bitrate action $a_t \in \mathcal{A}$ from a discrete bitrate ladder, where $\mathcal{A}$ is ordered from the lowest to the highest bitrate and $r(a_t)$ denotes the corresponding bitrate.

Let $S_t(a_t)$ denote the size of chunk $t$ under action $a_t$, and let $c_t$ denote the realized Starlink throughput during the chunk download. The download time is
\begin{equation}
    d_t(a_t)=\frac{8S_t(a_t)}{c_t}.
    \label{eq:download-time}
\end{equation}
The factor 8 converts chunk size from bytes to bits. Thus, a larger selected representation $S_t(a_t)$ or a lower Starlink throughput $c_t$ directly increases the download time and consumes more playback buffer.
Rebuffering occurs when the download time exceeds the available playback buffer:
\begin{equation}
    \rho_t = \max(d_t(a_t)-b_t,0).
    \label{eq:rebuffer}
\end{equation}
If $d_t(a_t)\leq b_t$, the chunk is downloaded before the current buffer is exhausted and $\rho_t=0$. Otherwise, the excess download time $d_t(a_t)-b_t$ is counted as rebuffering.
After the download, the buffer is updated as
\begin{equation}
    b_{t+1}=\min\{B_{\max},\max(b_t-d_t(a_t),0)+\Delta\},
    \label{eq:buffer-update}
\end{equation}
The inner maximum describes the remaining buffer after downloading. The newly downloaded chunk then adds $\Delta$ seconds of playable video, and the outer minimum enforces the maximum buffer size $B_{\max}$.

\subsection{QoE and Severe-Risk Metrics}
We evaluate an ABR policy at the session level by measuring both its accumulated QoE and its severe rebuffering risk. This session-level view captures the impact of rare Starlink throughput drops that may dominate user experience even when the expected QoE remains high.

Following common ABR formulations, the per-chunk QoE reward is
\begin{equation}
    q_t =
    \frac{r(a_t)}{1000}
    - \mu \rho_t
    - \eta \frac{|r(a_t)-r(a_{t-1})|}{1000},
    \label{eq:qoe}
\end{equation}
The first term rewards video quality, with the division by 1000 used to place the bitrate reward on the usual Mbps scale. The second term penalizes stall duration with weight $\mu$, and the third term penalizes bitrate switching with weight $\eta$. Therefore, a high-bitrate action is beneficial only when it does not create excessive rebuffering or quality oscillation. For one streaming session with $T$ chunks, we denote the session-level QoE reward $Q$ and cumulative rebuffering $R$ by
\begin{equation}
    Q=\sum_{t=1}^{T}q_t,
    \quad
    R=\sum_{t=1}^{T}\rho_t.
    \label{eq:session-qoe-rebuf}
\end{equation}

Let $\tau$ denote a random streaming session generated by a policy $\pi$ under the Starlink throughput process. The expected QoE objective is
\begin{equation}
    \mathcal{J}_Q(\pi)
    =
    \mathbb{E}_{\tau\sim \pi}[Q(\tau)] .
    \label{eq:expected-qoe}
\end{equation}
Maximizing this expectation alone can hide rare but severe session-level stalls. We therefore define a tail-rebuffering risk at confidence level $\beta$:
\begin{equation}
    \mathcal{R}_{\beta}(\pi)
    =
    \mathbb{E}_{\tau\sim \pi}
    \left[
    R(\tau)
    \mid
    R(\tau)\geq \mathrm{VaR}_{\beta}(R)
    \right],
    \label{eq:tail-risk}
\end{equation}
where $\mathrm{VaR}_{\beta}(R)$ is the $\beta$-quantile of the session-level rebuffering random variable. We also define the probability of a severe-stall session under a stall threshold $\rho_0$:
\begin{equation}
    \mathcal{S}_{\rho_0}(\pi)
    =
    \Pr_{\tau\sim \pi}\left(R(\tau)>\rho_0\right).
    \label{eq:severe-stall-prob}
\end{equation}
The QoE--severe-risk view in this paper is therefore to preserve a large $\mathcal{J}_Q(\pi)$ while reducing $\mathcal{R}_{\beta}(\pi)$ and $\mathcal{S}_{\rho_0}(\pi)$. The concrete finite-test estimators used in the experiments are given in Section~V. For training-time risk shaping, we use CVaR as the tail-risk measure. Let $X$ denote a rebuffering loss random variable induced by a policy. The CVaR at confidence level $\alpha$ is
\begin{equation}
    \mathrm{CVaR}_{\alpha}(X)
    =
    \mathbb{E}\left[X \mid X \geq \mathrm{VaR}_{\alpha}(X)\right],
    \label{eq:cvar}
\end{equation}
Here $\mathrm{VaR}_{\alpha}(X)$ is the $\alpha$-quantile of $X$, and CVaR averages the losses no smaller than this quantile.

\subsection{Safe-Capacity Action Feasibility}
We define a predicted feasible-action set that connects the safe-capacity estimate, the current playback buffer, and the bitrate-action feasibility before execution.

Let $\hat{c}_t$ be a predicted safe-capacity lower bound exposed to the ABR controller for the next chunk download. Given a buffer guard margin $g$, an action is predicted to be feasible if its estimated download time does not exceed the guarded buffer:
\begin{equation}
    \mathcal{F}_t(\hat{c}_t)
    =
    \left\{
    a \in \mathcal{A}
    \mid
    \frac{8S_t(a)}{\hat{c}_t}
    \leq b_t-g
    \right\}.
    \label{eq:feasible-set}
\end{equation}
The set $\mathcal{F}_t(\hat{c}_t)$ is the predicted safe action set. It keeps only those bitrates whose estimated download time under $\hat{c}_t$ fits within the guarded buffer $b_t-g$. If $b_t\leq g$, the guard leaves no positive safe download budget, and the auditor falls back to the lowest bitrate action. This feasibility set connects throughput prediction to ABR control: a capacity overestimate can incorrectly include an aggressive bitrate in $\mathcal{F}_t(\hat{c}_t)$, especially when the chunk is large or the buffer is low.

For analysis, we define a decision violation using the realized capacity:
\begin{equation}
    v_t(a)=
    \mathbf{1}
    \left\{
    \frac{8S_t(a)}{c_t}
    >
    b_t-g
    \right\}.
    \label{eq:decision-violation}
\end{equation}
Here, $\mathbf{1}\{\cdot\}$ denotes an indicator that returns one when the condition inside the braces is true and zero otherwise. Thus, $v_t(a)=1$ means that the actual download time of action $a$ exceeds the guarded buffer, while $v_t(a)=0$ means that the action satisfies the guarded-buffer constraint under the realized Starlink throughput. This metric is an ex-post safety check: it reveals whether an action that may appear feasible under prediction would actually be unsafe after execution.

\section{SafeSABR Design}

SafeSABR implements two complementary risk-control layers around a learned ABR policy. The training-time layer constructs a policy that starts from a high-QoE behavior-cloning prior and is then calibrated through risk-calibrated RL fine-tuning, instantiated by CVaR-PPO. The deployment-time layer uses a BG-CFQS safe-capacity estimate to audit the policy-requested bitrate before execution.

\begin{figure*}[!t]
  \centering
  \includegraphics[width=0.98\textwidth]{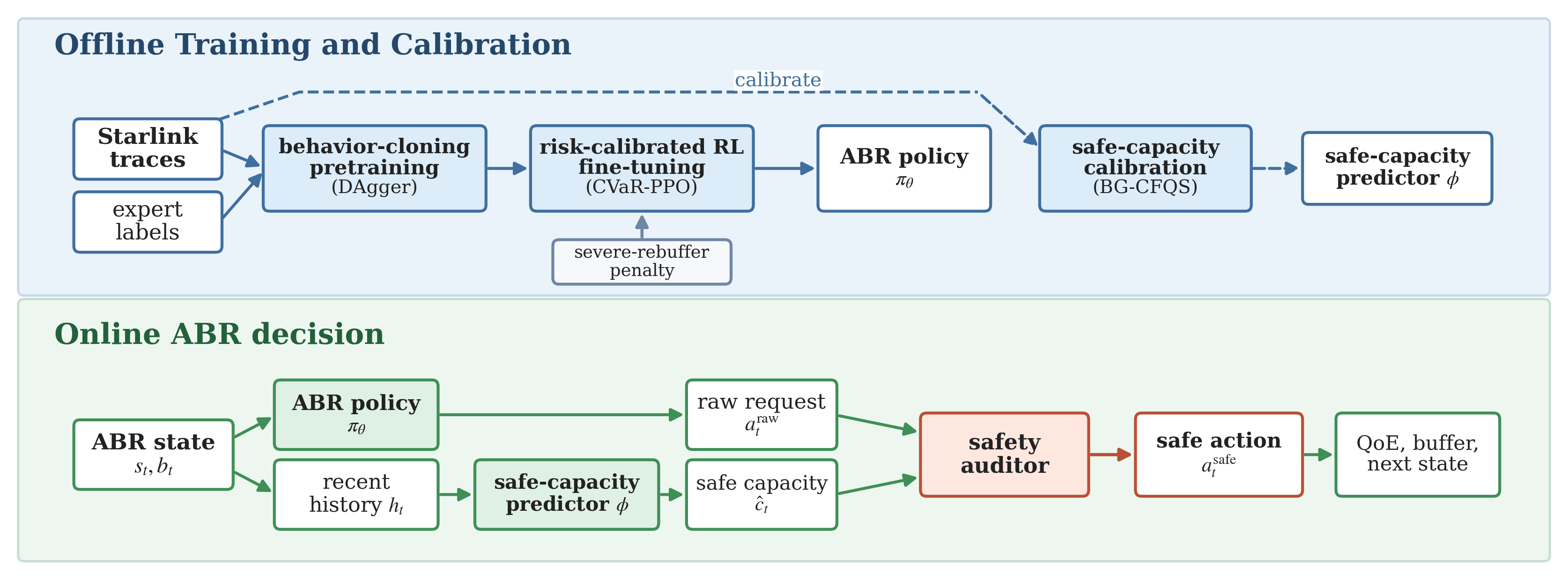}
  \caption{Framework of SafeSABR. The offline part constructs a high-QoE prior through behavior-cloning pretraining, applies risk-calibrated RL fine-tuning with CVaR-PPO, and calibrates a decision-aware safe-capacity predictor $\phi$ with BG-CFQS, while the online part uses $\phi$ to audit the policy-requested action before execution.}
  \label{fig:safesabr-framework}
\end{figure*}

\subsection{Framework Structure}
SafeSABR integrates risk-calibrated policy learning and decision-aware runtime auditing through the offline-to-online pipeline shown in Fig.~\ref{fig:safesabr-framework}. The offline part produces two artifacts: an ABR policy $\pi_\theta$ and a safe-capacity predictor $\phi$.

The policy-training path uses Starlink traces and expert labels to learn high-QoE bitrate behavior through behavior-cloning pretraining. It then performs risk-calibrated RL fine-tuning with CVaR-PPO, which penalizes severe rebuffering episodes and moves the policy toward a lower-risk action distribution while retaining the bitrate-selection capability learned from the expert. In parallel, the safe-capacity calibration path uses Starlink traces to calibrate BG-CFQS and outputs a predictor $\phi$ that provides the safe capacity used by the online auditor.

During online ABR decision making, the current ABR state is sent to the policy to obtain a raw bitrate request $a_t^{\mathrm{raw}}$, while the recent history $h_t$ is sent to the predictor to obtain $\hat{c}_t$. The safety auditor combines these two outputs at the action level: it keeps the raw request if it satisfies the feasible-action condition in \eqref{eq:feasible-set}, and otherwise downgrades it to the highest feasible lower bitrate. The executed action then determines the download time, QoE, buffer update, and next ABR state.

\subsection{Behavior-Cloning Pretraining}
SafeSABR first uses behavior cloning to learn a high-QoE initial policy from a reference ABR expert. This pretrained policy avoids random high-bitrate exploration and provides a useful ABR prior for the subsequent CVaR-PPO update.

Let $\pi_{\theta}(a|s)$ denote the probability that the learned policy with parameter $\theta$ selects action $a$ in state $s$, and let $\pi_E$ denote the expert policy. In our implementation, behavior-cloning pretraining uses DAgger-style data aggregation~\cite{ross2011reduction}: states visited by the current policy are aggregated into an imitation dataset $\mathcal{D}_{\mathrm{IL}}$, and the expert provides the corresponding action label $a^E_t=\pi_E(s_t)$.

The pretrained policy is obtained by minimizing the categorical cross-entropy imitation loss
\begin{equation}
    \mathcal{L}_{\mathrm{IL}}(\theta)
    =
    -\frac{1}{|\mathcal{D}_{\mathrm{IL}}|}
    \sum_{(s_t,a^E_t)\in \mathcal{D}_{\mathrm{IL}}}
    \sum_{a\in\mathcal{A}}
    y_t(a)\log \pi_{\theta}(a|s_t),
    \label{eq:il-loss}
\end{equation}
Here, $\mathcal{D}_{\mathrm{IL}}$ denotes the aggregated imitation dataset, whose elements are state-expert-action pairs $(s_t,a^E_t)$, and $|\mathcal{D}_{\mathrm{IL}}|$ is the number of such pairs. The term $y_t(a)=\mathbf{1}\{a=a^E_t\}$ is the one-hot expert label over the bitrate ladder $\mathcal{A}$. With this one-hot label, the cross-entropy loss is equivalent to the negative log-probability assigned to the expert action. Minimizing $\mathcal{L}_{\mathrm{IL}}$ therefore makes the learned policy imitate the expert on the states collected in $\mathcal{D}_{\mathrm{IL}}$. 

Algorithm~\ref{alg:bc-pretraining} summarizes the behavior-cloning pretraining procedure. Following Comyco-style imitation learning~\cite{huang2019comyco,huang2020quality}, the reference expert is implemented as a beam-search/MPC policy that uses future throughput over a finite horizon to generate high-QoE action labels for pretraining.

\begin{algorithm}[!t]
\caption{Behavior-Cloning Pretraining}
\label{alg:bc-pretraining}
\begin{algorithmic}[1]
\STATE \textbf{Input:} initial policy $\pi_\theta$, expert policy $\pi_E$, ABR simulator
\STATE \textbf{Input:} pretraining iterations $N_{\mathrm{pre}}$, rollout steps $T_{\mathrm{pre}}$, epochs $E_{\mathrm{pre}}$
\STATE Initialize imitation dataset $\mathcal{D}_{\mathrm{IL}}\leftarrow\emptyset$.
\FOR{$n=1,\ldots,N_{\mathrm{pre}}$}
    \STATE Roll out $\pi_\theta$ in the simulator for $T_{\mathrm{pre}}$ steps and collect visited states.
    \STATE Query $\pi_E$ for expert action labels on the collected states.
    \STATE Aggregate the labeled pairs into $\mathcal{D}_{\mathrm{IL}}$.
    \FOR{$e=1,\ldots,E_{\mathrm{pre}}$}
        \STATE Update $\theta$ by minimizing $\mathcal{L}_{\mathrm{IL}}(\theta)$ in \eqref{eq:il-loss}.
    \ENDFOR
\ENDFOR
\STATE \textbf{Output:} pretrained high-QoE ABR policy $\pi_\theta$
\end{algorithmic}
\end{algorithm}

\subsection{Risk-Calibrated RL Fine-Tuning}
Behavior cloning learns a high-QoE initial policy, but it does not explicitly control the severe-session rebuffering tail. SafeSABR therefore performs risk-calibrated RL fine-tuning, instantiated by combining proximal policy optimization (PPO)~\cite{schulman2017proximal} with a CVaR tail-risk penalty~\cite{rockafellar2000optimization,chow2015risksensitive,chow2018riskconstrained}. We refer to this instantiation as CVaR-PPO, so that severe rebuffering episodes influence the policy update more directly.

For a rollout batch $\mathcal{B}$ containing multiple trace episodes, let $R_i$ be the cumulative rebuffering of episode $i$ and let $|\mathcal{B}|$ be the number of episodes in the batch. The empirical CVaR penalty is estimated as
\begin{equation}
    \widehat{\mathrm{CVaR}}_{\alpha}(R)
    =
    \xi_{\alpha}
    +
    \frac{1}{(1-\alpha)|\mathcal{B}|}
    \sum_{i\in\mathcal{B}}
    [R_i-\xi_{\alpha}]_{+},
    \label{eq:cvar-estimator}
\end{equation}
Here $\xi_{\alpha}$ is the empirical $\alpha$-quantile of episode rebuffering within the rollout batch, and $[R_i-\xi_{\alpha}]_{+}$ measures how far episode $i$ lies above this tail threshold. Episodes below the threshold have zero excess term, while high-rebuffer episodes increase the penalty. The fine-tuning objective is
\begin{equation}
    J(\theta)
    =
    \mathbb{E}_{\pi_{\theta}}\left[\sum_{t=1}^{T} q_t\right]
    -
    \lambda \, \widehat{\mathrm{CVaR}}_{\alpha}(R),
    \label{eq:cvar-objective}
\end{equation}
The first term maximizes expected QoE over rollouts generated by $\pi_{\theta}$, while the second term subtracts a penalty for tail rebuffering. The weight $\lambda$ controls this tradeoff: larger values make the policy less willing to gain average quality through actions that create severe stalls on a small fraction of traces. In implementation, the risk-calibrated fine-tuning stage uses PPO-style policy updates~\cite{schulman2017proximal} with the CVaR tail penalty above. Algorithm~\ref{alg:cvar-ppo} summarizes the procedure.

\begin{algorithm}[!t]
\caption{Risk-Calibrated RL Fine-Tuning with CVaR-PPO}
\label{alg:cvar-ppo}
\begin{algorithmic}[1]
\STATE \textbf{Input:} pretrained policy $\pi_\theta$, ABR simulator, risk level $\alpha$, weight $\lambda$
\STATE \textbf{Input:} fine-tuning iterations $N_{\mathrm{RL}}$, rollout episodes per batch $M_{\mathrm{RL}}$
\FOR{$n=1,\ldots,N_{\mathrm{RL}}$}
    \STATE Roll out $\pi_\theta$ for $M_{\mathrm{RL}}$ trace episodes.
    \STATE Compute per-chunk rewards $\{q_t\}$ and episode rebuffering $\{R_i\}$.
    \STATE Estimate $\widehat{\mathrm{CVaR}}_{\alpha}(R)$ by \eqref{eq:cvar-estimator}.
    \STATE Update $\pi_\theta$ using PPO with the objective in \eqref{eq:cvar-objective}.
\ENDFOR
\STATE \textbf{Output:} risk-calibrated ABR policy $\pi_\theta$
\end{algorithmic}
\end{algorithm}

\subsection{Decision-Aware Safe-Capacity Predictor Interface}
Risk-calibrated RL fine-tuning reduces unsafe action tendencies, but online auditing still requires a capacity value for the next chunk. SafeSABR uses a decision-aware safe-capacity predictor interface: the predictor exposes one scalar $\hat{c}_t$ that is consumed by the feasibility set in \eqref{eq:feasible-set}, rather than being evaluated only as a standalone throughput forecast:
\begin{equation}
    \hat{c}_t = \phi(h_t),
    \label{eq:safe-capacity-interface}
\end{equation}
The predictor $\phi$ maps the available history, and Starlink-side features $h_t$ to one scalar capacity value used by the auditor~\cite{xie2026riskawaresafethroughputforecasting,liu2025starnet}. A point predictor can set $\hat{c}_t$ to its point forecast, while a lower-bound or risk-aware predictor returns a more conservative value intended to reduce harmful overestimation.

The predictor is not judged only by symmetric point error. Its output is consumed by the feasible action set in \eqref{eq:feasible-set}; therefore, an overestimated $\hat{c}_t$ can make an unsafe high-bitrate action appear feasible. Algorithm~\ref{alg:safe-capacity} summarizes the generic construction of the safe-capacity predictor used by SafeSABR. The candidate set can include a point predictor and calibrated lower-bound predictors, and the final deployable SafeSABR uses BG-CFQS~\cite{xie2026riskawaresafethroughputforecasting} as the risk-aware safe-capacity predictor.

\begin{algorithm}[!t]
\caption{Decision-Aware Safe-Capacity Predictor Construction}
\label{alg:safe-capacity}
\begin{algorithmic}[1]
\STATE \textbf{Input:} Starlink trace history, predictor candidates $\Phi$, calibration traces
\STATE \textbf{Input:} ABR decision samples and feasibility rule in \eqref{eq:feasible-set}
\FOR{each candidate predictor $\phi_j\in\Phi$}
    \STATE Train or calibrate $\phi_j$ on Starlink history.
    \STATE Generate safe-capacity estimates $\hat{c}_t^{(j)}=\phi_j(h_t)$ on calibration traces.
    \STATE Build feasible sets $\mathcal{F}_t(\hat{c}_t^{(j)})$ according to \eqref{eq:feasible-set}.
    \STATE Evaluate decision violations for admitted actions according to \eqref{eq:decision-violation}.
    \STATE Evaluate downstream ABR rebuffer risk under this predictor.
\ENDFOR
\STATE Select the predictor $\phi$ according to the desired QoE--severe-risk operating point.
\STATE \textbf{Output:} safe-capacity predictor $\phi$
\end{algorithmic}
\end{algorithm}

\begin{figure}[!t]
  \centering
  \includegraphics[width=\linewidth]{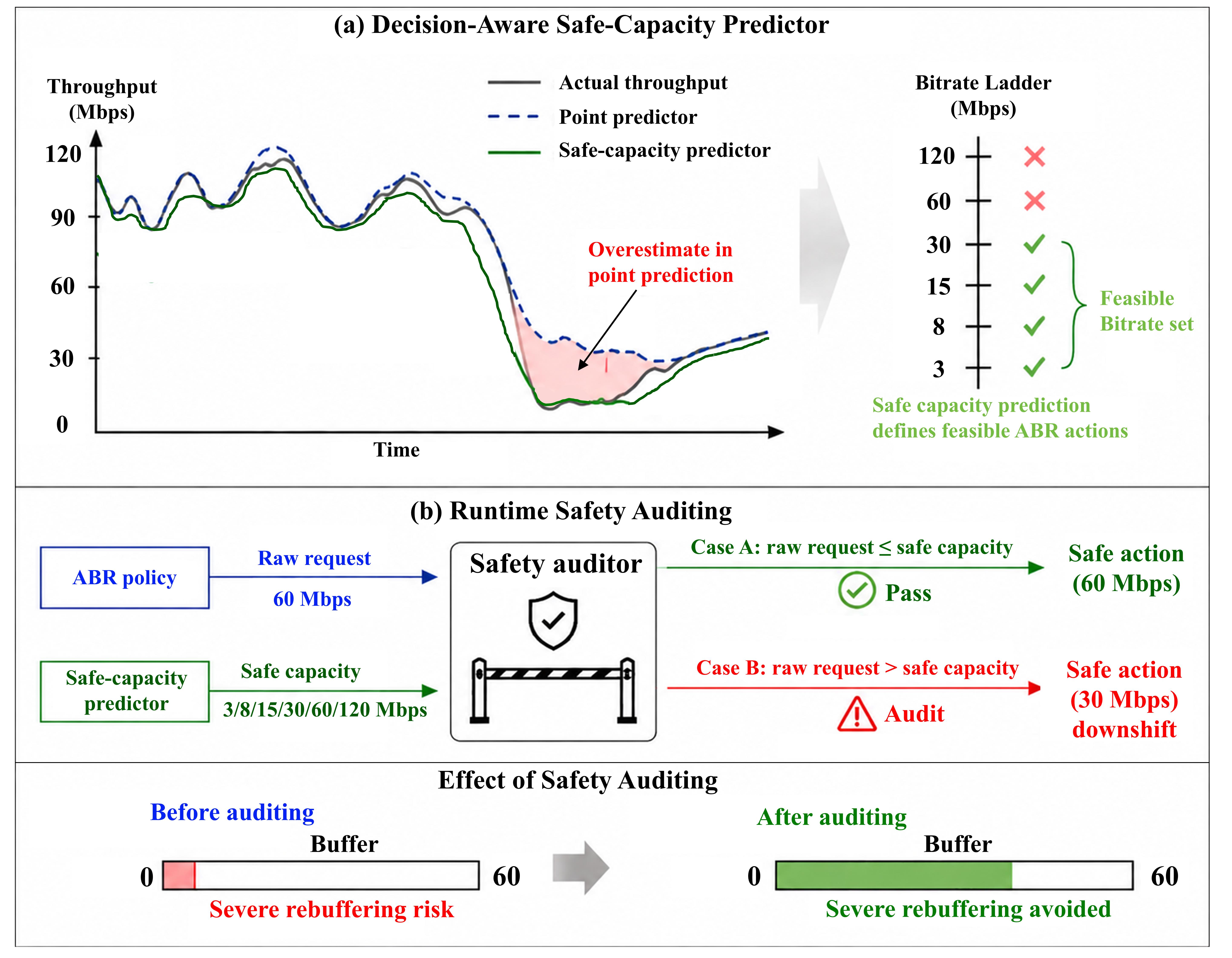}
  \caption{Decision-aware safe-capacity prediction and runtime safety auditing.}
  \label{fig:auditor-principle}
\end{figure}

\subsection{Runtime Safety Auditing}
Even after risk-calibrated RL fine-tuning, a learned policy can request an unsafe bitrate under abrupt Starlink drops. Runtime safety auditing is the deployment-time risk-control layer of SafeSABR: it preserves feasible policy actions and changes only requests that violate the predicted buffer-safety constraint.

At deployment time, the learned policy first requests a raw action $a_t^{\mathrm{raw}}$. Since $\mathcal{A}$ is ordered by bitrate, the relation $a\leq a_t^{\mathrm{raw}}$ means that action $a$ corresponds to a bitrate no higher than the requested bitrate. The safety auditor then computes the largest action no higher than the request that satisfies the predicted feasibility condition:
\begin{equation}
    a_t^{\mathrm{safe}}
    =
    \begin{cases}
    \max \left\{
    a \in \mathcal{F}_t(\hat{c}_t)
    \mid
    a \leq a_t^{\mathrm{raw}}
    \right\},
    &
    \text{if the set is nonempty},\\
    a_{\min}, & \text{otherwise},
    \end{cases}
    \label{eq:projection}
\end{equation}
If the raw request is already feasible, the maximum feasible action no higher than the request is the request itself, so the learned decision is preserved. If the request is unsafe, the auditor searches downward on the ordered bitrate ladder and executes the highest lower bitrate that satisfies the buffer guard. If no action satisfies the guard, the lowest bitrate $a_{\min}$ is used as a fallback. Thus, the auditor acts as a decision-level safety check rather than a blanket bitrate reduction rule.

We also record the intervention indicator
\begin{equation}
    m_t=\mathbf{1}\{a_t^{\mathrm{safe}}\neq a_t^{\mathrm{raw}}\},
    \label{eq:intervention}
\end{equation}
The indicator $m_t$ equals one only when the auditor changes the policy request. Therefore, the average of $m_t$ over chunks measures how often deployment-time safety correction is needed. A lower intervention rate under comparable tail risk indicates better consistency between the trained policy and the safe-capacity constraint.

Algorithm~\ref{alg:safety-auditing} summarizes the runtime safety-auditing procedure. Fig.~\ref{fig:auditor-principle} illustrates how the safe-capacity interface and runtime auditor work together. The predictor converts volatile Starlink throughput forecasts into a safe capacity and the corresponding feasible bitrate set. The auditor then preserves a feasible policy request, but downshifts an unsafe request to the highest feasible lower bitrate, protecting the playback buffer from severe depletion.

\begin{algorithm}[!t]
\caption{Runtime Safety Auditing}
\label{alg:safety-auditing}
\begin{algorithmic}[1]
\STATE \textbf{Input:} policy $\pi_\theta$, predictor $\phi$, bitrate ladder $\mathcal{A}$
\STATE \textbf{Input:} ABR state $s_t$, recent history $h_t$, feasibility rule in \eqref{eq:feasible-set}
\STATE Select raw action $a_t^{\mathrm{raw}}\leftarrow\arg\max_{a\in\mathcal{A}}\pi_\theta(a|s_t)$.
\STATE Predict safe capacity $\hat{c}_t\leftarrow\phi(h_t)$.
\STATE Construct $\mathcal{F}_t(\hat{c}_t)$ according to \eqref{eq:feasible-set}.
\STATE Let $\mathcal{C}_t\leftarrow\{a\in\mathcal{F}_t(\hat{c}_t)\mid a\leq a_t^{\mathrm{raw}}\}$.
\IF{$\mathcal{C}_t$ is nonempty}
    \STATE Set $a_t^{\mathrm{safe}}\leftarrow\max \mathcal{C}_t$.
\ELSE
    \STATE Set $a_t^{\mathrm{safe}}\leftarrow a_{\min}$, where $a_{\min}$ is the lowest bitrate action.
\ENDIF
\STATE Set $m_t\leftarrow\mathbf{1}\{a_t^{\mathrm{safe}}\neq a_t^{\mathrm{raw}}\}$.
\STATE \textbf{Output:} executed action $a_t^{\mathrm{safe}}$ and intervention indicator $m_t$
\end{algorithmic}
\end{algorithm}

\section{Performance Evaluation}
This section evaluates whether SafeSABR improves the QoE--severe-risk operating point of high-bitrate ABR over real Starlink traces. After introducing the experimental setting, the main comparison and operating-point plot examine whether SafeSABR reduces severe-session rebuffering while preserving high QoE. The regional and handover-heavy experiments then test whether this behavior remains visible across different Starlink trace groups and under mobility-induced stress. A hard-trace case study further illustrates how runtime auditing changes bitrate decisions inside a difficult session. Finally, the predictor-auditor analysis, staged ablation, and sensitivity study isolate the roles of safe-capacity forecasting, risk-calibrated RL fine-tuning, runtime auditing, and their key parameters.

\subsection{Experimental Setting}
\label{sec:experimental-setting}
\subsubsection{Datasets and ABR Task}
We evaluate SafeSABR using the real Starlink measurement datasets~\cite{liu2025starnet}. The dataset records downlink throughput from Starlink terminals across multiple regions and captures the satellite-network dynamics, including frequent serving-satellite changes and handovers. These properties make the dataset suitable for evaluating whether ABR algorithms can maintain high QoE without creating severe-session stalls under rapid LEO access-link fluctuations. 

The dataset includes regional measurements from Chicago, USA; OSN, Germany; and Victoria, Canada. We process these measurements into three regional datasets and denote them as \emph{US}, \emph{OSN}, and \emph{VIC}, respectively. For ABR evaluation, each continuous throughput sequence is converted into SABR replay traces, where each row records time and measured throughput at 1-s granularity. The ABR simulator then downloads video chunks over these measured Starlink throughput traces. Table~\ref{tab:abr-dataset-summary} summarizes the StarNet measurement source, processed throughput samples, handover statistics, and trace-set split. The trace-set split is reported in the format of train/calibration/test trace counts.

\begin{table*}[!t]
\centering
\caption{Summary of Starlink ABR datasets.}
\label{tab:abr-dataset-summary}
\begin{tabular}{lcccc}
\toprule
\textbf{Item} & \textbf{US} & \textbf{OSN} & \textbf{VIC} & \textbf{Total} \\
\midrule
Location & Chicago, USA & OSN, Germany & Victoria, Canada & -- \\
Processed period & 2024-04-26--2024-05-28 & 2024-07-13--2024-07-31 & 2024-07-11--2024-07-28 & -- \\
Throughput samples & 1,123,832 & 613,295 & 145,053 & 1,882,180 \\
Trace minutes & 41,252 & 10,221 & 2,417 & 53,890 \\
Satellite handovers & 86,808 & 26,782 & 7,257 & 120,847 \\
Trace-set split & 169/36/36 & 91/19/20 & 109/24/23 & 369/79/79 \\
\bottomrule
\end{tabular}
\end{table*}

The training split is used to train behavior-cloning and RL policies. The calibration split is used to select safe-capacity operating points and predictor-auditor parameters. The test split is held out for final reporting. To stress ABR decisions under high-capacity but volatile satellite access, we use a 4K/8K-style high-bitrate ladder and chunk configuration, as summarized later in Table~\ref{tab:safesabr-exp-config}.

\subsubsection{Baselines and Method Variants}
We compare SafeSABR with representative online, prediction-assisted, and learned ABR methods. The non-learned online baselines include BOLA~\cite{spiteri2020bola} and RobustMPC~\cite{yin2015control}. The prediction-assisted online baselines include Lumos-MPC~\cite{lv2022lumos,lv2024accurate} and StarNet-MPC~\cite{liu2025starnet}, which combine throughput prediction with the same MPC decision logic. BOLA makes bitrate decisions from buffer occupancy, RobustMPC uses history-based robust throughput estimation with a five-chunk planning horizon, Lumos-MPC feeds a Lumos-style decision-tree throughput predictor into MPC, and StarNet-MPC feeds StarNet point throughput predictions into MPC. The learned baselines include Pensieve~\cite{mao2017neural}, Comyco~\cite{huang2019comyco}, and SABR~\cite{luo2025sabr}. Pensieve represents RL-based neural ABR, Comyco represents imitation-learning-based ABR, and SABR denotes the behavior-cloning-pretrained policy with vanilla PPO fine-tuning~\cite{schulman2017proximal}.

For mechanism analysis, we compare several safe-capacity predictor variants. StarNet-point denotes the point-throughput forecast produced by the StarNet predictor~\cite{liu2025starnet}. StarNet-LB denotes a calibrated lower-bound (LB) version of StarNet, where the point forecast is converted into a conservative safe-capacity estimate on the calibration split. Xgboost-LB uses an XGBoost~\cite{chen2016xgboost} regression model with the same lower-bound calibration interface. BG-CFQS is the risk-aware safe throughput forecasting method proposed in~\cite{xie2026riskawaresafethroughputforecasting}; it provides calibrated lower-bound capacity estimates for Starlink control and is used as the default safe-capacity predictor in SafeSABR. These variants separate point prediction, lower-bound prediction, risk-aware forecasting, and runtime auditing.

\subsubsection{Implementation Details and Hyperparameters}
All learned policies use the same Starlink high-bitrate setting unless otherwise specified. Behavior cloning uses a beam-search expert~\cite{huang2019comyco,huang2020quality}, and PPO is implemented with Stable-Baselines3~\cite{raffin2021stable} using a multi-layer perceptron (MLP) policy. The CVaR penalty is used only during training; all reported rewards are computed with the original QoE reward in \eqref{eq:qoe}. Safe-capacity predictors are calibrated only on the held-out calibration traces before being evaluated on the test traces. Table~\ref{tab:safesabr-exp-config} summarizes the main experimental configuration and hyperparameters.

\begin{table*}[!t]
\centering
\caption{Implementation and hyperparameter configuration.}
\label{tab:safesabr-exp-config}
\begin{tabular}{p{0.24\textwidth}p{0.68\textwidth}}
\toprule
\textbf{Item} & \textbf{Value} \\
\midrule
\multicolumn{2}{l}{\emph{ABR simulator}} \\
Bitrate ladder & 3, 8, 15, 30, 60, 120 Mbps \\
Video setting & 48 chunks, 4 s per chunk, mild variable-bitrate chunk-size variation \\
Playback buffer & Maximum buffer 60 s \\
QoE reward weights & Rebuffer penalty $\mu=40$, smoothness penalty $\eta=1$ \\
\midrule
\multicolumn{2}{l}{\emph{Behavior-cloning pretraining}} \\
Expert and data aggregation & Beam-search expert, 15 DAgger iterations, 2000 rollout steps per iteration \\
Behavior-cloning optimizer & Adam, learning rate $10^{-3}$, batch size 128, 5 epochs per DAgger iteration \\
Behavior-cloning loss & Expert-action negative log-likelihood, equivalent to categorical cross-entropy; entropy coefficient 0 \\
\midrule
\multicolumn{2}{l}{\emph{Risk-calibrated RL fine-tuning}} \\
PPO implementation & Stable-Baselines3 PPO with MLP policy, 4 parallel environments, 50k training steps \\
PPO rollout/update & $n_{\mathrm{steps}}=512$ per environment, batch size 64, 10 epochs per update \\
PPO optimization & Learning rate $3\times10^{-4}$, $\gamma=0.99$, generalized advantage estimation $\lambda=0.95$, clip range 0.2 \\
PPO regularization & Entropy coefficient 0, value-loss coefficient 0.5, max gradient norm 0.5 \\
Normalization & Reward normalization enabled with clipping 10; observation normalization disabled \\
CVaR-PPO instantiation & Mode \texttt{cvar\_rebuf}, $\alpha=0.90$, penalty weight $\lambda=20$, budget 0, window 512 \\
\midrule
\multicolumn{2}{l}{\emph{Predictor and implementation}} \\
Safe-capacity predictor & Input length 75 s, prediction horizon 15 s, calibrated on the calibration split \\
\bottomrule
\end{tabular}
\end{table*}

\subsubsection{Evaluation Metrics}
All final results are reported on the held-out test traces using finite-sample QoE and severe-risk metrics. For each evaluated session $i\in\mathcal{I}$, we compute its session QoE $Q_i$ and cumulative rebuffering $R_i$ according to \eqref{eq:session-qoe-rebuf}. Average QoE and mean rebuffering are reported as
\begin{equation}
    \bar{Q}
    =
    \frac{1}{|\mathcal{I}|}
    \sum_{i\in\mathcal{I}} Q_i,
    \quad
    \bar{R}
    =
    \frac{1}{|\mathcal{I}|}
    \sum_{i\in\mathcal{I}} R_i .
    \label{eq:test-average-metrics}
\end{equation}
To estimate the tail-rebuffering risk in \eqref{eq:tail-risk} with $\beta=0.95$, we report the average cumulative rebuffering of the worst 5\% sessions:
\begin{equation}
    R_{\mathrm{worst5}}
    =
    \frac{1}{K}
    \sum_{i\in \mathrm{Top}_{K}(\{R_j\}_{j\in\mathcal{I}})}
    R_i,
    \quad
    K=\left\lceil0.05|\mathcal{I}|\right\rceil .
    \label{eq:session-tail}
\end{equation}
Here $\mathrm{Top}_{K}(\cdot)$ returns the indices of the $K$ evaluated sessions with the largest cumulative rebuffering values. To estimate the severe-stall probability in \eqref{eq:severe-stall-prob} with $\rho_0=10$ s, we report the fraction of sessions with more than 10 s cumulative rebuffering:
\begin{equation}
    S_{>10}
    =
    \frac{1}{|\mathcal{I}|}
    \sum_{i\in\mathcal{I}}
    \mathbf{1}\{R_i>10\}.
    \label{eq:severe-session-ratio}
\end{equation}

In experiments involving the runtime auditor, we also report the audit intervention rate, i.e., the average value of $m_t$ in \eqref{eq:intervention}, to quantify how often the auditor changes the policy request. For the predictor-auditor analysis, let $\mathcal{D}_{\mathrm{dec}}$ be the decision-evaluation samples and let $a_t^{\phi}$ be the action admitted by predictor $\phi$. The decision-violation rate is
\begin{equation}
    V_{\mathrm{dec}}
    =
    \frac{1}{|\mathcal{D}_{\mathrm{dec}}|}
    \sum_{t\in\mathcal{D}_{\mathrm{dec}}}
    v_t(a_t^{\phi}).
    \label{eq:decision-violation-rate}
\end{equation}
To focus on harmful positive prediction errors in difficult link states, we also report the high-risk overestimation rate over low-throughput samples:
\begin{equation}
    \mathrm{OverRate}_{\mathrm{HR}}
    =
    \frac{1}{|\mathcal{D}_{\mathrm{HR}}|}
    \sum_{t\in\mathcal{D}_{\mathrm{HR}}}
    \mathbf{1}\{\hat{c}_t>c_t\},
    \label{eq:highrisk-overrate}
\end{equation}
where $\mathcal{D}_{\mathrm{HR}}$ contains samples whose realized throughput belongs to the lowest 30\% of the evaluation set.

\subsection{Comparison with Advanced ABR Methods}
This experiment provides the comparison against representative ABR methods and tests the central question of this paper: whether SafeSABR can reduce severe session-level rebuffering without collapsing average QoE. Table~\ref{tab:main-results} reports the result as a QoE--severe-risk tradeoff using the session-level metrics defined above. BOLA and RobustMPC are used as conservative low-risk references, while Lumos-MPC, StarNet-MPC, Pensieve, Comyco, and SABR represent high-QoE prediction-assisted or learned baselines. The QoE Cost column is computed only for high-QoE baselines and indicates the relative QoE decrease of SafeSABR with respect to each baseline.

\begin{table*}[!t]
\centering
\caption{QoE--severe-risk tradeoff on Starlink traces.}
\label{tab:main-results}
\begin{tabular}{llccccc}
\toprule
\textbf{Group} & \textbf{Method} & \textbf{QoE} & \textbf{Mean Rebuf (s)} & \textbf{Worst-5\% Rebuf (s)} & \textbf{Session $>10$s (\%)} & \textbf{QoE Cost} \\
\midrule
Conservative ref. & BOLA & 4108.66 & 1.42 & 14.12 & 2.5\% & -- \\
Conservative ref. & RobustMPC & 4568.16 & 1.74 & 16.33 & 3.8\% & -- \\
High-QoE baseline & Lumos-MPC & 4660.07 & 8.07 & 63.83 & 22.8\% & 0.6\% \\
High-QoE baseline & StarNet-MPC & 4684.98 & 5.92 & 40.40 & 21.5\% & 1.2\% \\
High-QoE baseline & Pensieve & 4644.89 & 20.49 & 134.40 & 35.0\% & 0.3\% \\
High-QoE baseline & Comyco & 4732.19 & 7.38 & 39.72 & 25.3\% & 2.1\% \\
High-QoE baseline & SABR & 4715.33 & 7.77 & 54.30 & 22.8\% & 1.8\% \\
Ours & SafeSABR & 4630.81 & 2.50 & 22.68 & 7.2\% & -- \\
\bottomrule
\end{tabular}
\end{table*}

The conservative references confirm that rebuffering risk can be reduced by choosing lower-quality operating points: BOLA and RobustMPC have low severe-stall ratios, but their QoE scores are 4108.66 and 4568.16. In contrast, high-QoE baselines expose much larger severe session tails. Lumos-MPC, StarNet-MPC, Pensieve, Comyco, and SABR reach QoE values between 4644.89 and 4732.19, but their worst-5\% session rebuffering ranges from 39.72 s to 134.40 s, and 21.5\%--35.0\% of their sessions exceed 10 s of cumulative rebuffering. SafeSABR keeps a comparable QoE level of 4630.81 while reducing mean rebuffering to 2.50 s, worst-5\% session rebuffering to 22.68 s, and the severe-stall session ratio to 7.2\%. Compared with SABR, SafeSABR reduces mean rebuffering from 7.77 s to 2.50 s and severe-stall sessions from 22.8\% to 7.2\%, with a 1.8\% QoE cost. Thus, the main value of SafeSABR is not to maximize average QoE, but to move high-QoE ABR toward a safer severe-risk operating point.

\subsection{QoE--Severe-Risk Tradeoff}
\begin{figure*}[!t]
  \centering
  \includegraphics[width=1\textwidth]{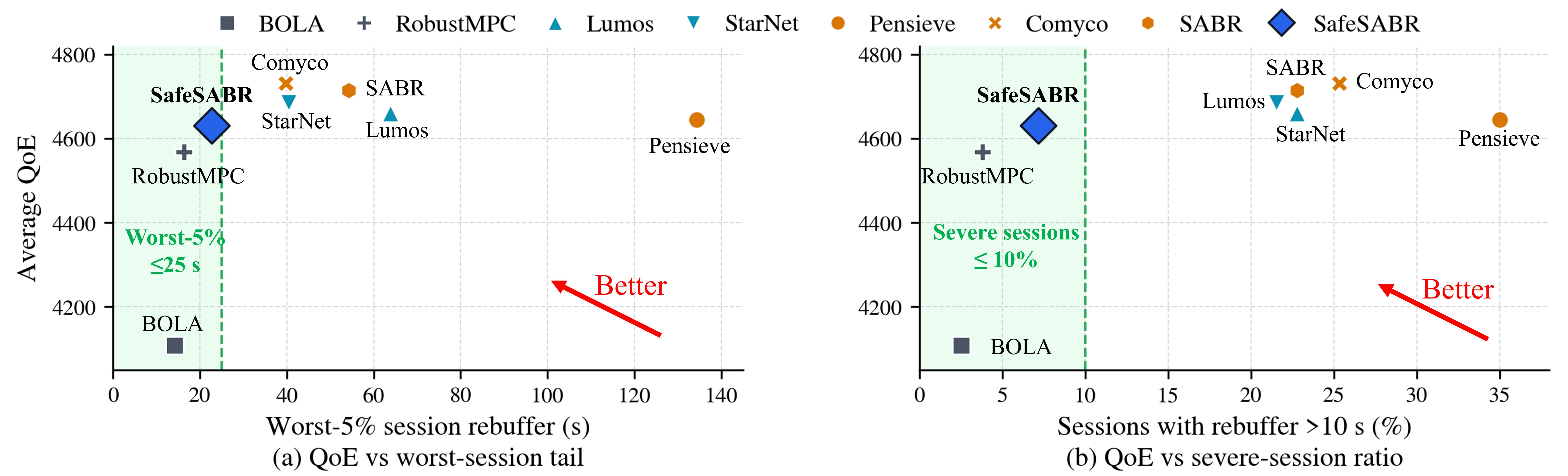}
  \caption{QoE--severe-risk operating points on Starlink traces.}
  \label{fig:qoe-severe-risk-tradeoff}
\end{figure*}

This experiment examines the operating position of each method in the QoE--severe-risk space. To make the tradeoff visible, Fig.~\ref{fig:qoe-severe-risk-tradeoff} plots average QoE against two complementary severe-risk estimates: worst-5\% session rebuffering and the fraction of sessions with more than 10 s cumulative rebuffering. The vertical dashed line in each panel marks the preferred low-risk side, and the red arrow indicates the better direction along the risk axis.

BOLA and RobustMPC remain on the preferred low-risk side, but they also appear in the lower-QoE region. Lumos-MPC, StarNet-MPC, Pensieve, Comyco, and SABR move upward in QoE, while most of their points fall to the right of the risk boundary, especially in the severe-session-ratio view. SafeSABR is not the topmost point in QoE, but it moves back toward the preferred side of both risk axes while staying close to the high-QoE group. This visual pattern supports the main claim of SafeSABR: it improves the QoE--severe-risk operating point rather than simply maximizing average QoE or conservatively lowering bitrate quality.

\subsection{Robustness Across Starlink Regions}
\begin{figure*}[!t]
  \centering
  \includegraphics[width=0.85\textwidth]{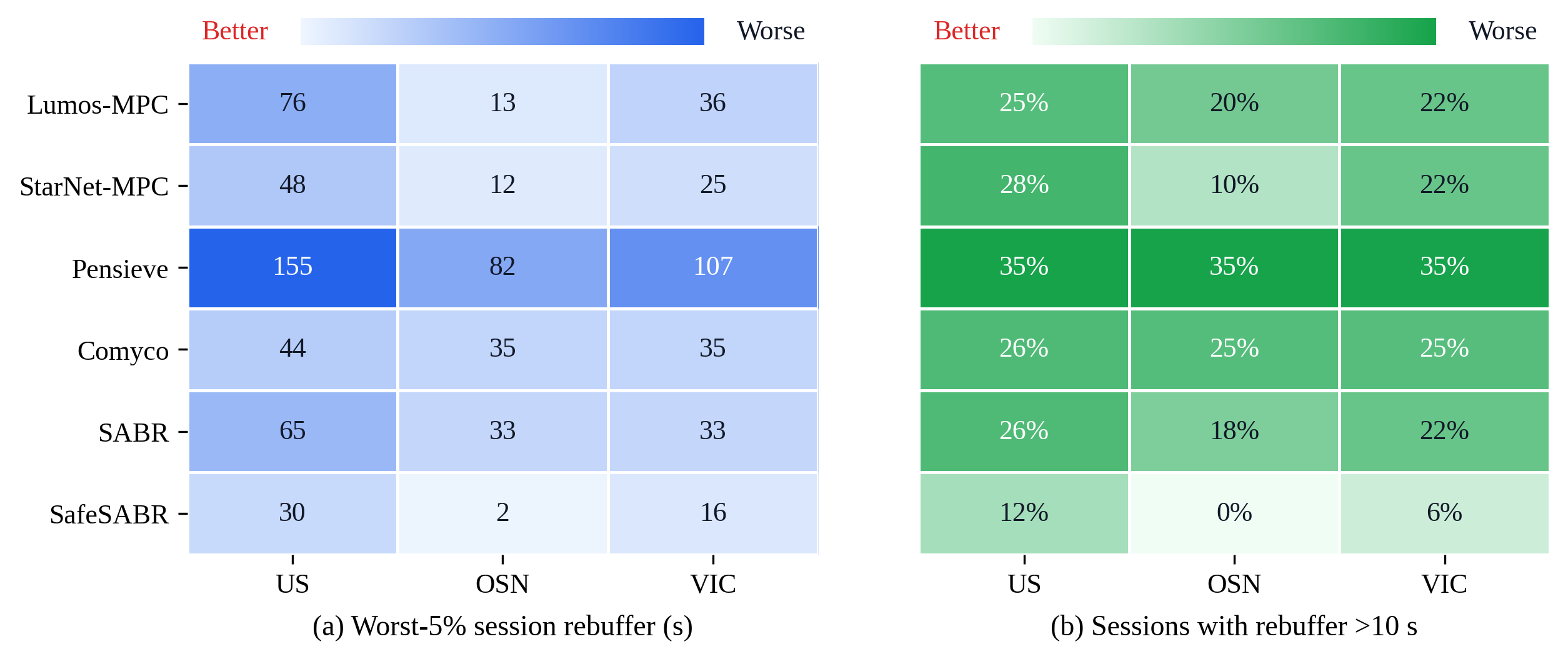}
  \caption{Robustness across Starlink regions.}
  \label{fig:region-robustness}
\end{figure*}

This experiment evaluates whether the QoE--severe-risk advantage of SafeSABR remains stable across different Starlink measurement regions. The US, OSN, and VIC trace groups exhibit different throughput distributions and handover patterns, providing a regional robustness test beyond the aggregate result. Fig.~\ref{fig:region-robustness} reports severe-session metrics separately for each region. The two heatmaps correspond to the worst-5\% session rebuffering and severe-stall session ratio, respectively. Darker cells indicate larger severe-session risk, so a robust method should remain lighter across regions rather than only performing well on the aggregate test set.

The US trace group is the hardest case for all high-QoE methods, but SafeSABR still keeps the worst-5\% session rebuffering at 29.5 s, compared with 48.1 s for StarNet-MPC, 43.5 s for Comyco, and 64.7 s for SABR. On OSN and VIC, SafeSABR further reduces worst-5\% session rebuffering to 2.1 s and 15.7 s, respectively. The severe-stall session ratio follows the same trend: SafeSABR has 12.0\%, 0.0\%, and 5.8\% severe-stall sessions on US, OSN, and VIC, lower than the high-QoE baselines in every region. This indicates that the SafeSABR gain is not tied to a single regional trace group.

\subsection{Stress Test on Handover-Heavy Traces}
This experiment evaluates SafeSABR under Starlink mobility stress by focusing on handover-heavy traces. The goal is to examine how frequently serving-satellite changes amplify severe-session rebuffering and whether the proposed risk-calibrated control remains effective in these difficult periods.

\begin{figure*}[!t]
  \centering
  \includegraphics[width=0.94\textwidth]{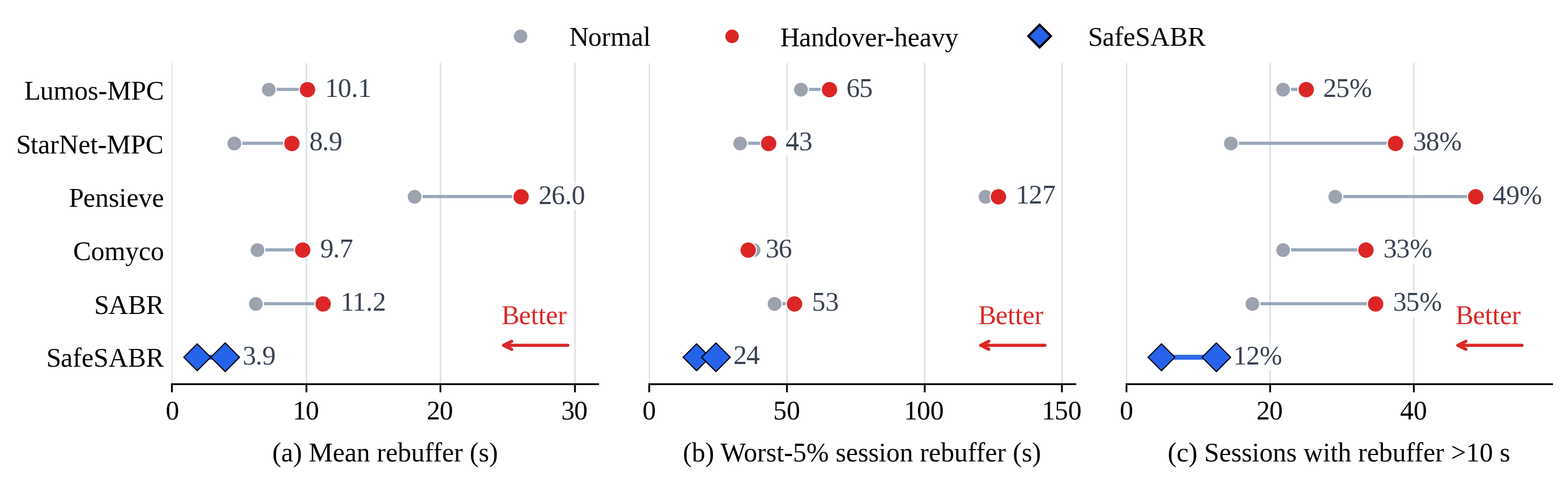}
  \caption{Stress test on handover-heavy Starlink traces.}
  \label{fig:handover-stress}
\end{figure*}

We construct the stress subset from the Starlink handover metadata. For each test trace, we count serving-satellite changes during the first 300 s of the measurement sequence, which covers the playback window used by the 48-chunk ABR evaluation. Within each StarNet region, the top 30\% traces by handover count are labeled handover-heavy, with large one-second throughput drops used only to break ties. This produces 24 handover-heavy traces, with 11 from US, 6 from OSN, and 7 from VIC. In Fig.~\ref{fig:handover-stress}, each line connects the normal and handover-heavy results of the same method, so a rightward shift indicates that handover-heavy periods amplify playback risk.

The handover-heavy traces contain 17.2 handovers on average during the 300-s window, compared with 14.6 for normal traces, and this stress subset increases playback risk for all high-QoE baselines. For example, StarNet-MPC mean rebuffering rises from 4.61 s to 8.92 s, and SABR rises from 6.25 s to 11.25 s. The severe-stall session ratio also increases from 14.5\% to 37.5\% for StarNet-MPC, from 17.6\% to 34.7\% for SABR, and from 4.8\% to 12.5\% for SafeSABR. On handover-heavy traces, SafeSABR keeps mean rebuffering at 3.95 s, worst-5\% session rebuffering at 24.22 s, and severe-stall sessions at 12.5\%. In contrast, StarNet-MPC, Comyco, SABR, and Pensieve reach 37.5\%, 33.3\%, 34.7\%, and 48.6\% severe-stall sessions, respectively. This shows that the SafeSABR advantage becomes more relevant under Starlink-specific handover stress.

\subsection{Hard-Trace Mechanism Case Study}
\begin{figure}[!t]
  \centering
  \includegraphics[width=\linewidth]{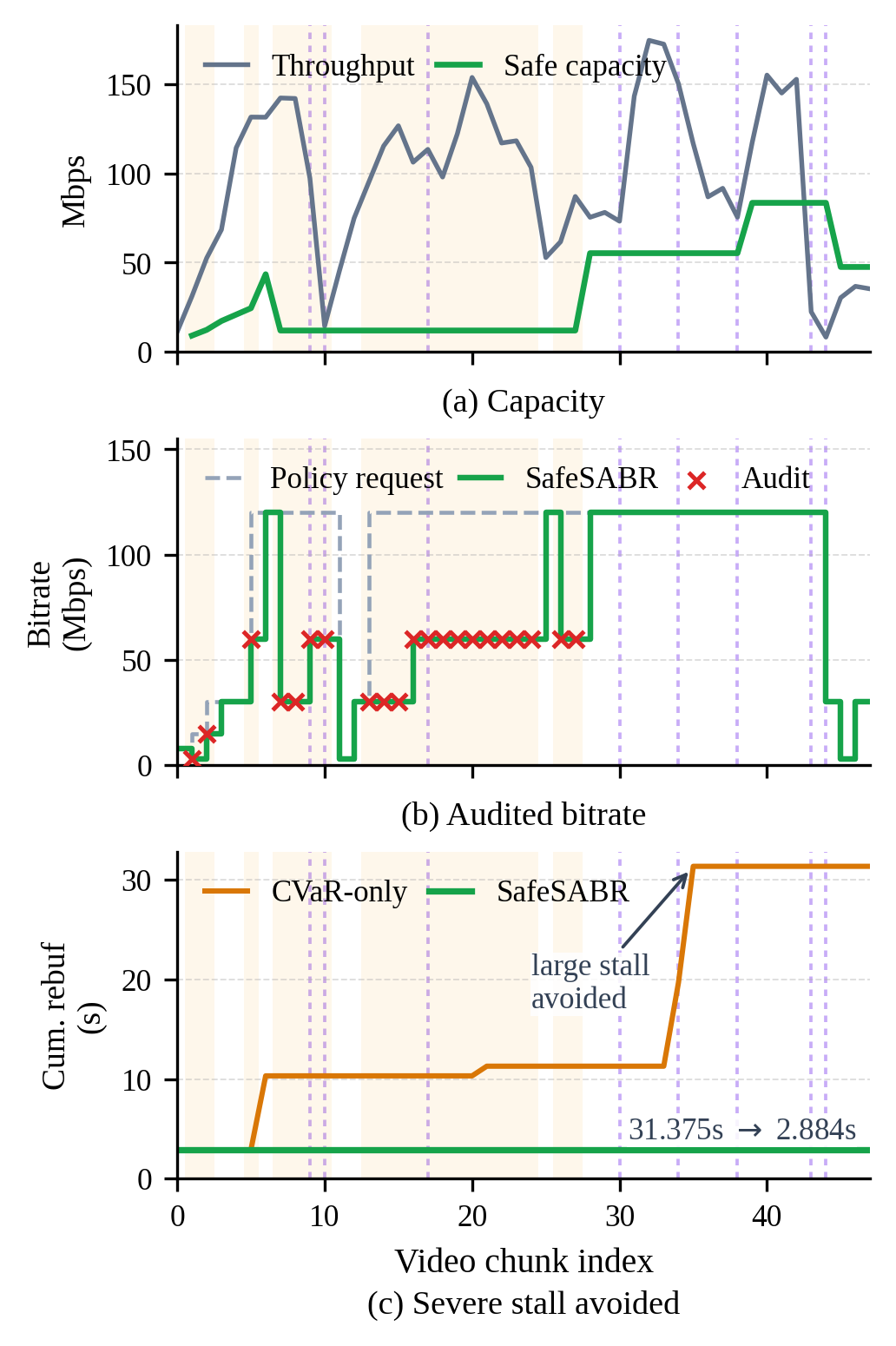}
  \caption{Mechanism case study on a representative hard Starlink trace.}
  \label{fig:mechanism-case}
\end{figure}

The aggregate results show that SafeSABR reduces severe-session stalls, but they do not show when the deployment-time safety layer changes a bitrate decision inside a session. Therefore, this case study inspects one hard Starlink trace in which the risk-calibrated policy still requests aggressive bitrates during throughput drops and frequent serving-satellite handovers. Fig.~\ref{fig:mechanism-case} summarizes the mechanism in three steps: panel (a) compares the measured throughput with the BG-CFQS safe-capacity estimate used by the auditor, panel (b) shows how the auditor maps policy-requested bitrates to SafeSABR-executed bitrates, and panel (c) compares cumulative rebuffering with and without runtime auditing.

Reading the three panels together shows that the auditor is not active throughout the session. It changes the requested action only when the requested bitrate remains high while the BG-CFQS safe-capacity estimate falls. These selective corrections reduce the executed bitrate before the severe-stall region shown in panel (c), preventing the large cumulative rebuffering jump that appears when SafeSABR is used without runtime auditing. On this trace, cumulative rebuffering decreases from 31.375 s to 2.884 s, while the total QoE increases from 1389.00 to 2566.63 because avoiding the long stall outweighs the temporary bitrate reductions. This example explains the mechanism behind the aggregate results: SafeSABR does not obtain safety by uniformly lowering bitrate quality, but by auditing high-risk decisions under difficult Starlink states.

\subsection{Predictor and Safety Auditor Analysis}
This experiment evaluates whether safe-capacity prediction provides useful decision information for the runtime auditor. Instead of testing standalone throughput prediction accuracy, it measures whether different predictor interfaces reduce unsafe bitrate admissions and downstream severe-session rebuffering.

Table~\ref{tab:predictor-decision} compares four predictor interfaces under the same ABR control pipeline. To isolate the effect of safe-capacity prediction, all rows use the same SafeSABR policy before runtime auditing and the same auditing rule, while only the predictor input to the auditor is changed. The two decision-side metric columns measure predictor risk through decision violation and high-risk overestimation, while the remaining metric columns report downstream ABR rebuffering and audit frequency. A useful safe-capacity predictor should reduce decision violations and high-risk overestimation, and this reduction should translate into lower severe-session playback risk without requiring excessive audit intervention.

\begin{table*}[!t]
\centering
\caption{Predictor and safety-auditor analysis under the same pre-audit SafeSABR policy.}
\label{tab:predictor-decision}
\begin{tabular}{lccccc}
\toprule
\textbf{Predictor} & \textbf{Dec. Viol.} & \boldmath$\mathrm{OverRate}_{\mathrm{HR}}$\unboldmath & \textbf{Mean Rebuf (s)} & \textbf{Worst-5\% Rebuf (s)} & \textbf{Audit Rate} \\
\midrule
StarNet-point & 26.6\% & 75.5\% & 3.10 & 26.16 & 3.1\% \\
StarNet-LB & 21.6\% & 70.3\% & 2.85 & 24.24 & 3.5\% \\
Xgboost-LB & 21.4\% & 73.6\% & 2.95 & 24.98 & 3.2\% \\
BG-CFQS & \textbf{20.5\%} & \textbf{69.3\%} & \textbf{2.50} & \textbf{22.68} & 4.3\% \\
\bottomrule
\end{tabular}
\end{table*}

StarNet-point has the highest decision violation rate and high-risk overestimation rate, showing that point prediction is not sufficient for safety-critical ABR decisions. StarNet-LB lowers both rates by exposing the auditor to a conservative capacity estimate. Xgboost-LB has a similar average decision violation rate to StarNet-LB, but its larger high-risk overestimation rate leaves a larger severe-session tail. BG-CFQS gives the lowest decision violation, high-risk overestimation, mean rebuffering, and worst-5\% session rebuffering among the compared predictors. Its audit rate is only 4.3\%, and the QoE scores of the predictor variants remain within a narrow range of 4619.18--4630.81. Therefore, the gain is not obtained by crudely lowering bitrate quality; it comes from using a safer predictor-auditor interface for ABR decisions.

\subsection{Ablation Study of SafeSABR}
This experiment quantifies the contribution of each SafeSABR design stage to QoE preservation and severe-risk reduction. We construct controlled configurations by enabling or disabling behavior-cloning pretraining, risk-calibrated RL fine-tuning, and the Safety Auditor. In Table~\ref{tab:safesabr-ablation}, checkmarks indicate which components are enabled in each configuration.

\begin{table*}[!t]
\centering
\caption{Ablation of SafeSABR design stages using session-level severe-risk metrics.}
\label{tab:safesabr-ablation}
\begin{tabular}{ccc@{\hspace{1.2em}}cccc}
\toprule
\multicolumn{3}{c}{\textbf{Components}} & \multicolumn{4}{c}{\textbf{Metrics}} \\
\cmidrule(lr){1-3}\cmidrule(lr){4-7}
\textbf{Behavior Cloning} & \textbf{Risk-calibrated Fine-tuning} & \textbf{Safety Auditor} & \textbf{QoE} & \textbf{Mean Rebuf (s)} & \textbf{Worst-5\% Rebuf (s)} & \textbf{Session $>10$s (\%)} \\
\midrule
\checkmark & -- & -- & \textbf{4738.33} & 8.88 & 49.32 & 28.7\% \\
\checkmark & \checkmark & -- & 4683.64 & 4.74 & 30.14 & 17.3\% \\
\checkmark & -- & \checkmark & 4657.80 & 3.15 & 29.67 & 9.7\% \\
\checkmark & \checkmark & \checkmark & 4630.81 & \textbf{2.50} & \textbf{22.68} & \textbf{7.2\%} \\
\bottomrule
\end{tabular}
\end{table*}

Behavior-cloning gives the highest QoE, but it leaves 28.7\% of sessions with more than 10 s of cumulative rebuffering. Adding risk-calibrated RL fine-tuning reduces mean rebuffering from 8.88 s to 4.74 s and lowers the severe-stall session ratio from 28.7\% to 17.3\%, showing the effect of training-time risk calibration before any runtime correction. Adding the Safety Auditor without risk-calibrated fine-tuning also reduces severe stalls, but the full SafeSABR configuration gives the lowest mean rebuffering, worst-5\% session rebuffering, and severe-stall session ratio. This result supports the intended division of labor: behavior cloning provides a high-QoE prior, risk-calibrated RL fine-tuning performs training-time calibration, and the runtime auditor corrects residual unsafe bitrate requests.

\subsection{Sensitivity to Risk and Auditing Parameters}
This experiment analyzes the two risk-control parameters that determine the SafeSABR operating point: the CVaR penalty weight in risk-calibrated RL fine-tuning and the margin used by the runtime auditor. Table~\ref{tab:sensitivity} separates the analysis into two blocks. The first block varies the CVaR penalty weight without runtime auditing, isolating the effect of training-time risk calibration. The second block fixes $\lambda=20$ and varies the BG-CFQS auditor margin, isolating the effect of deployment-time action auditing. Each block is compared with its corresponding no-risk-control or no-audit row.

\begin{table*}[!t]
\centering
\caption{Sensitivity to risk-training and runtime auditing parameters.}
\label{tab:sensitivity}
\begin{tabular}{llccccc}
\toprule
\textbf{Parameter} & \textbf{Setting} & \textbf{QoE} & \textbf{Mean Rebuf (s)} & \textbf{Worst-5\% Rebuf (s)} & \textbf{Session $>10$s (\%)} & \textbf{Audit Rate} \\
\midrule
CVaR weight & $\lambda=0$ & 4715.33 & 7.77 & 54.30 & 22.8\% & -- \\
CVaR weight & $\lambda=10$ & 4710.56 & 6.30 & 35.80 & 20.3\% & -- \\
CVaR weight & $\lambda=20$ & 4683.64 & 4.74 & 30.14 & 17.3\% & -- \\
CVaR weight & $\lambda=40$ & 4674.43 & 5.03 & 34.76 & 16.5\% & -- \\
\midrule
Auditor margin & No audit & 4683.64 & 4.74 & 30.14 & 17.3\% & 0.0\% \\
Auditor margin & $m=0.90$ & 4630.81 & 2.50 & 22.68 & 7.2\% & 4.3\% \\
Auditor margin & $m=0.95$ & 4638.40 & 2.54 & 22.57 & 8.0\% & 4.0\% \\
Auditor margin & $m=1.00$ & 4638.62 & 2.61 & 22.41 & 7.6\% & 3.8\% \\
\bottomrule
\end{tabular}
\end{table*}

Without runtime auditing, increasing the CVaR penalty from $\lambda=0$ to $\lambda=20$ reduces worst-5\% session rebuffering from 54.30 s to 30.14 s and lowers the severe-stall session ratio from 22.8\% to 17.3\%, with a small QoE decrease. A larger penalty $\lambda=40$ further lowers the severe-stall ratio to 16.5\%, but it does not improve the mean or worst-5\% rebuffering over $\lambda=20$. After fixing $\lambda=20$, the BG-CFQS auditor gives similar severe-risk reductions across margins $m=0.90$, $0.95$, and $1.00$. All three audited settings reduce worst-5\% session rebuffering from 30.14 s to about 22--23 s and reduce severe-stall sessions from 17.3\% to 8.0\% or lower, while auditing only 3.8\%--4.3\% of chunks. The final setting $m=0.90$ is selected because it gives the lowest mean rebuffering and severe-stall session ratio among the audited settings, not because it is an isolated optimum.

\section{Conclusion}
This paper presented SafeSABR, a risk-calibrated prediction-control framework for adaptive bitrate streaming over Starlink networks. We formulated high-bitrate Starlink ABR as a QoE--severe-risk tradeoff problem and used session-level metrics to expose severe rebuffering that can be hidden by average QoE. SafeSABR combines behavior-cloning pretraining for a high-QoE ABR prior, risk-calibrated RL fine-tuning with CVaR-PPO for reducing severe-tail action tendencies, and a BG-CFQS-driven runtime safety auditor for correcting unsafe bitrate requests at deployment time. Experiments on real Starlink network traces show that SafeSABR reduces severe-stall sessions from 22.8\% for SABR to 7.2\%, with a 1.8\% QoE cost. Predictor-auditor, ablation, and sensitivity analyses further support the complementary roles of risk-calibrated RL fine-tuning, decision-aware safe-capacity prediction, and runtime auditing. These results support using risk-calibrated learned control, safe-capacity auditing, and QoE--severe-risk evaluation together for high-bitrate ABR over volatile LEO satellite links. Future work will explore how large language model-based artificial intelligence agents can assist Starlink ABR systems by interpreting network context, coordinating prediction-control modules, and adapting safety policies across heterogeneous application scenarios.

\bibliographystyle{IEEEtran}
\bibliography{ref}

\vfill

\end{document}